# Picosecond-precision optical time transfer in free space using flexible binary offset carrier modulation

**HONGLEI YANG,**[1,2,3,*] **HAIFENG WANG,**[1,3] **XUEYUN WANG,**[1] **HANG YI,**[1] **WENZHE YANG,**[1] **HONGBO WANG,**[1] **AND SHENGKANG ZHANG**[1]

[1] *Science and Technology on Metrology and Calibration Laboratory, Beijing Institute of Radio Metrology and Measurement, Beijing 100854, China*
[2] *State Key Laboratory of Precision Measurement Technology & Instruments, Department of Precision Instrument, Tsinghua University, Beijing 100084, China*
[3] *These authors equally contributed to this work.*
*\*yhlpc@163.com*

**Abstract:** Free-space optical time transfer that features high precision and flexibility will act a crucial role in near-future ground-to-satellite/inter-satellite clock networks and outdoor timing services. Here we propose a free-space optical flexible-binary-offset-carrier-modulated (FlexBOC-modulated) time transfer method. The utilized FlexBOC modulation could yield a comparative precision, although its occupied bandwidth is tremendously reduced by at least 97.5% compared to optical binary phase modulation. Meanwhile, the adoption of optical techniques eliminates the multi-path effect that is major limit in the current microwave satellite time transfer system. What's more, the time interval measurement avoids a continuous link that may be routinely broken by physical obstructions. For verification, a time transfer experiment with our home-built system between two sites separated by a 30-m free-space path outside the laboratory was conducted. Over a 15 h period, the time deviation is 2.3 ps in a 1-s averaging time, and averages down to 1.0 ps until ~60 s. The fractional frequency instability exhibits $4.0 \times 10^{-12}$ at a gate time of 1 s, and approaches to $2.6 \times 10^{-15}$ at 10000 s.

## 1. Introduction

Nowadays, positioning, navigation and timing (PNT) capabilities and technologies have been promoting rapid developments in diverse fields, such as civilian activity, financial trade, transportation management, electric power dispatch, logistics, etc. Among these technologies, timing service lays a solid foundation of the others. In recent years, the performances of optical atomic clocks have continued to improve and have reached unprecedented stability and accuracy at a level of $10^{-18}$ [1-3], Therefore, time and frequency transfer has become the bottleneck of timing.

Conventional time and frequency transfer methods are based on free-space microwave communication through geostationary (GEO) satellites, typically referred to as two-way satellite time and frequency transfer (TWSTFT) , and global navigation satellite systems (GNSS) [4]. By simultaneously sending and receiving microwave signals between a pair of exchange stations through an intermediate GEO satellite, common disturbance along the path could be almost cancelled out. The time transfer instability reaches a few parts in $10^{13}$ at 1 s [5]. The GNSS method is more simple, but at the cost of real-time operation [6,7].

More precision methods have been explored in optical domain, where could achieve better accuracy, higher stability and boarder bandwidth. In present, practical high-precision optical time and frequency transfer is conducted via fiber-link networks. With microwave modulation, time stamp information and RF frequency standard could be simultaneously encoded into optical signal, and then transferred along urban fiber link to remote side at picosecond level limited by electronics [8]. The frequency uncertainty could reach several parts in $10^{15}$ at 1 s via phase detection and compensation [8,9]. Coherent optical frequency transfers via hundreds of kilometers fiber links between separated state-of-the-art narrow-linewidth lasers using fiber noise cancellation could achieve a residual instability of $10^{-15}$ at 1 s [10-14]. Due to its flexibility, free-space optical time and frequency is more and more attractive. Microwave transfer through a free-space optical communication channel has shown a uncertainty better than $1\times10^{-13}$ in 1 s over ~100 m distance [15]. Free-space coherent optical frequency transfer has exhibited a frequency uncertainty better than several parts in $10^{16}$ at 1 s over a 18-km horizontal path on the ground [16], and owns a potential uncertainty below the $4\times10^{-16}$ at 1 s over a vertical link to Low Earth Orbit [17]. Based on this technique, the Kepler system aims to reach sub-femtosecond-level synchronization across the whole constellation, including Low Earth Orbit and Middle Earth Orbit satellites [18]. Recently, comb-based optical two-way time and frequency transfer over a 2-km free-space link has demonstrated femtosecond-level synchronization with time-of-flight method [19-22], and further achieved attosecond-level transfer via carrier-phase technique [23]. This performance allows precise intercomparisons between optical lattice clocks. Nevertheless, such a high precision is not always required in general applications. Free-space optical two-way time and frequency with digital coherent communication has reached sub-picosecond timing precision across a 4-km turbulent range [24]. Besides, several ground-to-satellite projects, such as time transfer by laser link (T2L2) [25] and European laser timing (ELT) [26], have realized picosecond level transfer.

Here we propose an alternative picosecond-level precision free-space optical time transfer using flexible binary offset carrier (FlexBOC, or previous FBOC) modulation. Thus it could be simply established on a practical TWSTFT system. Though the occupied bandwidth is tremendously reduced compared to binary phase modulation, the utilized FlexBOC modulation could achieve a comparative precision via comprehensive analysis of the pseudorandom noise (PN) code and sub-carrier within FlexBOC signal, while the adoption of optical techniques eliminates the multi-path effect due to microwave reflections that is major limit in the current TWSTFT. Moreover, the time interval measurement avoids a continuous link that may be routinely broken by physical obstructions.

## 2. Principle

Due to effective suppression of time delay fluctuation along the link, two-way methodology is widely used in precision time transfer [4]. In a two-way transfer configuration, Site A and B send their waveforms at global timestamps: $T_{AT}$ and $T_{BT}$, respectively. The waveforms reciprocally propagate along a common path, and are received in the opposite sites at global timestamps: $T_{BR}$ and $T_{AR}$, respectively. As illustrated in Fig. 1(a), the time intervals between sending local waveform and receiving remote waveform in Site A and B, $T_A$ and $T_B$, could be described as

$$T_A = T_{AR} - T_{AT} = (T_{AR} - T_{BT}) + \Delta T, \qquad (1)$$

$$T_B = T_{BR} - T_{BT} = (T_{BR} - T_{AT}) - \Delta T, \qquad (2)$$

respectively, where $\Delta T$ is the clock difference between the Site A and B. However, the round-bracket terms in Eq. 1 and 2 are impossible for measurement. As shown in Fig. 1(b), these terms can be substituted as

$$T_{AR} - T_{BT} = \tau_{TXB} + \tau_{BA} + \tau_{RXA}, \qquad (3)$$

$$T_{BR} - T_{AT} = \tau_{TXA} + \tau_{AB} + \tau_{RXB}, \qquad (4)$$

where $\tau_{TXi}$ ($i$ = A, B) is the time delay from transmitter to optical terminator, $\tau_{RXi}$ ($i$ = A, B) is the time delay from optical terminator to receiver, $\tau_{AB}$ and $\tau_{BA}$ are time delays through the link along both forward and backward directions, respectively. According to Eq. 1-4, the clock time difference is computed by

$$\Delta T = \frac{1}{2}[(T_A - T_B) - (\tau_{BA} - \tau_{AB}) - (\tau_{TXB} + \tau_{RXA} - \tau_{TXA} - \tau_{RXB})], \qquad (5)$$

where $\tau_{AB}$ and $\tau_{BA}$ are identical in an assumption of the full reciprocity of the link. The other terms, ($\tau_{TXA}$-$\tau_{RXA}$) and ($\tau_{TXB}$-$\tau_{RXB}$), are fixed differential time delays and could be calibrated.

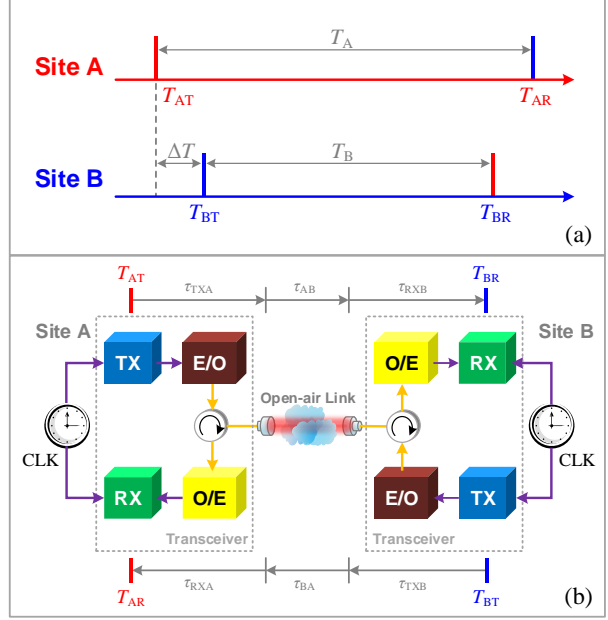

**Fig. 1.** Principle of optical two-way time transfer. (a) Timing diagram; (b) System architecture. TX, transmitter; RX, receiver; E/O, electronic-to-optical converter; O/E, optical-to-electronic converter; CLK, clock.

## 3. Experimental Setup and Module Design

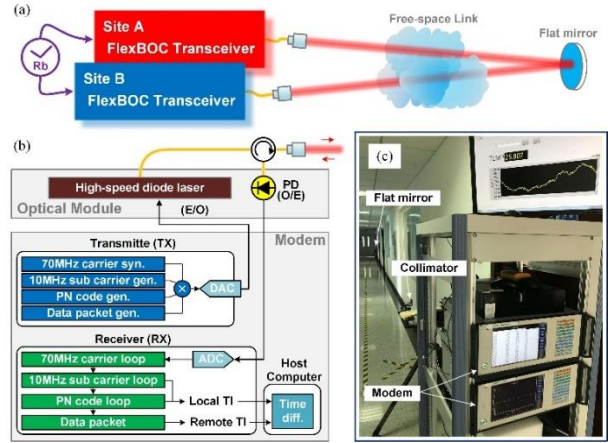

**Fig. 2.** (a) Schematic of the free-space optical time transfer system using FlexBOC modulation. (b) Top-level design of the FlexBOC transceiver. (c) Snapshot of the experimental system. syn, synthesizer; gen., generator; ADC, analog-to-digital converter; DAC, digital-to-analog converter; PD, photodiode; PN, pseudorandom noise; TI, Time interval; Time diff., Time difference.

According to the principle, a free-space optical time transfer system using FlexBOC modulation was established, as shown in Fig. 2. The transceivers in Sites A and B were co-located, but not necessarily, to share a Rubidium frequency standard for the convenience of verification. Each FlexBOC transceiver is in the identical design including an optical module and a modem, and fulfills both time interval measurement and data interaction in full duplex mode. A free-space optical communication link between both sites was then built via C-band laser lights encoded by FlexBOC modulation [27], and fully exploited the reciprocity of the atmosphere to suppress the air turbulence along the traversing path [4]. The differential length variations of fiber-coupled circulators caused by temperature excursion are reduced in a passive way.

### 3.1 Optical Module

In Fig. 3(a), the home-built optical module contains a laser diode and a photodiode to achieve E/O and O/E conversion, respectively. The diode laser that emits 4-mW optical light at ~1550 nm is capable to high-speed current modulation. Correspondingly, the photodiode is optimized to allow a broadband photodetection. Within the E/O part, the laser diode operates in constant-current mode to realize a broadband modulation. The operating current is regulated by a feedback loop. The modulation input that is coupled through the capacitor drives the output power of diode laser, and therefore encodes laser light. Within the O/E part, laser input is firstly converted to electronic signal by photodiode. Then, two stages of broadband amplifiers followed by signal conditioners boost the power of receiving signal. The modulation bandwidth of the optical module was tested by a microwave network analyzer and presented in Fig. 3(b). It shows that the optical module supports an analog modulation up to ~3.5 GHz, which benifits a high-fidelity E/O and O/E conversion of FlexBOC signal.

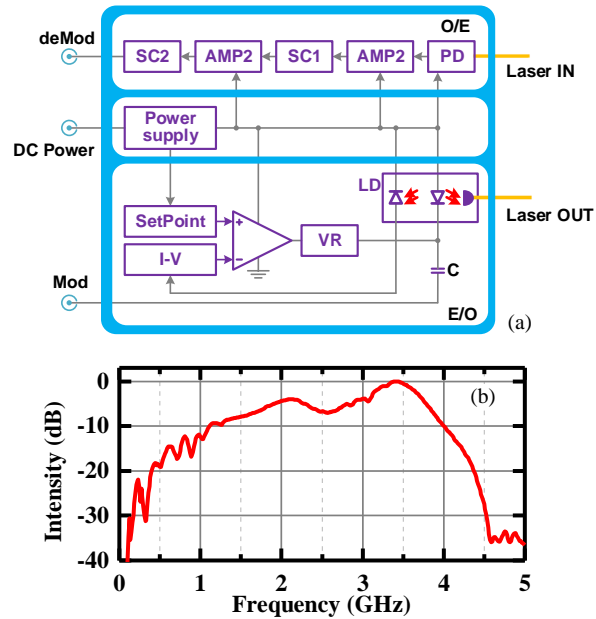

**Fig. 3.** (a) Functional architecture of optical module. (b) Bandwidth test. Mod, modulation input; deMod, demodulation output; I-V, current-to-voltage conversion; VR, voltage regulator; C, capacitor; LD, laser diode; PD, photodiode; AMP, amplifier; SC, signal conditioner.

### 3.2 FlexBOC Modem

Our homebuilt prototypes of FlexBOC modems are generally developed for TWSTFT [28]. The transmitter sends local FlexBOC signal, which contains the following four components. A 70-MHz intermediate frequency (IF) carrier is synthesized to shift the modulation from the optical carrier. A 10-MHz sub-carrier is added for fine time interval measurement, while a PN code is used for coarse time interval measurement. One PN code sequence contains 125 chips with 8 μs chip duration (125 kHz chip rate). Therefore, the PN code sequences repeat every 1 ms and enable clock time difference measurement at a maximum rate of 1 kHz. The measured time intervals and other interactive information are combined as digital codes for date exchange between individual sites. The four signal components are mixed together in time domain, and then used to modulate the driving current of the high-speed diode laser in the optical module.

As illustrated in Fig. 4(a), due to the sub-carrier within a FlexBOC signal, two sidebands are generated around the IF carrier with an offset of 10 MHz, namely at (IF-10) MHz and (IF+10) MHz. The introduced 125 kHz PN codes bring cascaded sidebands around the sub-carrier. The inset of Fig. 4(a) gives an expanded view of the sidebands. In the signal processing within the receiver, the high-order harmonics of the PN code are filtered out, while the remaining fundamental component is used for the phase analysis [28]. Thus, the typical effective bandwidth of the FlexBOC signal in our design is 500 kHz. We also simulated the spectrum of the BPSK signal that is adopted in a similar free-space optical two-way time synchronization system. Its 10-MHz chip rate occupied an effective bandwidth of 20 MHz [24], as depicted in Fig. 4(b). In the next section, we will show that our system could exhibit a comparative performance. To some extent, the proposed method benefits a dramatic release of occupied bandwidth in one link, typically at least 97.5 %.

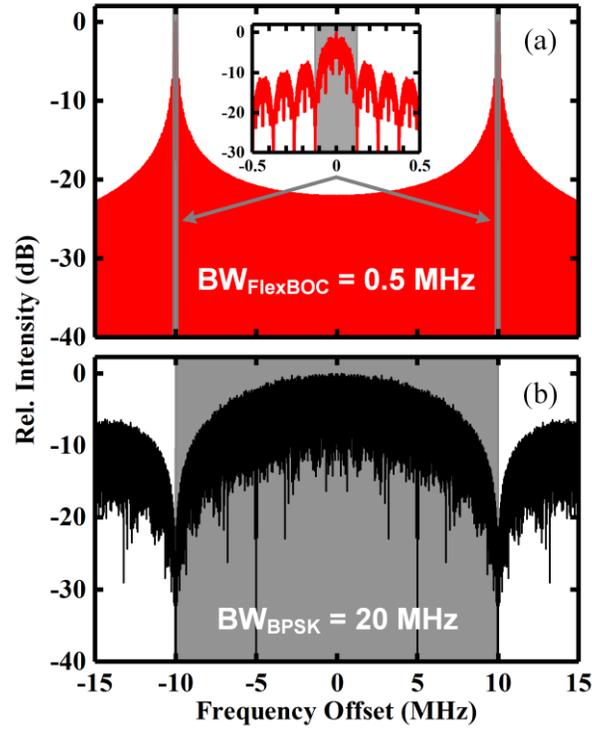

**Fig. 4.** Comparison between the spectra of (a) FlexBOC signal and (b) BPSK signal. The gray shadows mask their effective bandwidths

The receiver is used to find the time intervals between the events of sending local FlexBOC signal and receiving remote FlexBOC signal. The functional design has been described in detail in Reference [28]. In general, BPSK-based systems use autocorrelation method to realize time interval measurement, due to the single-peak waveform (according to the sampling rate of digitizer) [24]. However, resolving and tracking the autocorrelation of FlexBOC signal significantly burdens computational overhead and complicates processing algorithm, since multiple autocorrelation peaks are generated by the introduced sub-carrier, as shown in Fig. 5. Instead, we planted three digital phase-lock loops to extract the signal components, i.e., the 70-MHz IF carrier, 10-MHz square sub-carrier, PN code and interactive data, and to track PN code and sub-carrier in parallel [28,29]. Practically, the comprehensive time interval measurement with autocorrelation of PN codes and phase analysis of sub-carrier could reach a picosecond level precision, which is close to the performance of BPSK method. After the local time intervals are sent by transmitter to the remote site, the clock time difference can be found in both sites according to Eq. 5.

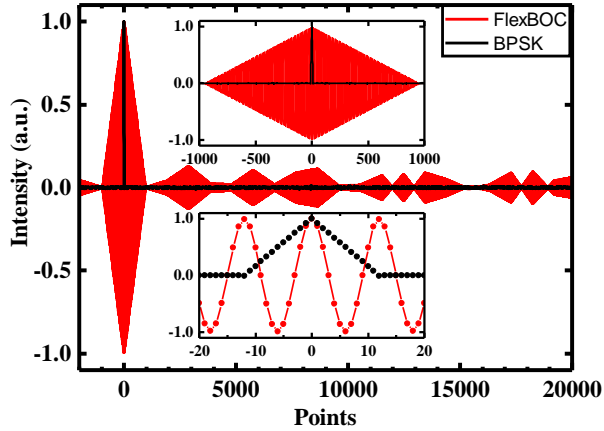

**Fig. 5.** Numerical simulation of the autocorrelation of FlexBOC signal (red line) and BPSK signal (black line). The top inset gives a clear view of the centerburst, and the bottom inset shows the peak of the autocorrelation.

## 4. Results

Figure 6 exhibits measured RF spectra of the FlexBOC signals in the modem. Although there is a ~40-dB loss between the transmitter and receiver separated by a 30-m free-space path, picosecond-level transfer still maintained for a long period. Such a great loss is mainly from the low power-receiving efficiency of the collimators. However, the received weak FlexBOC signal could be extracted due to the autocorrelation in the signal processing, even if it is submerged in noise. Hence, a transmission power of -40 dBm could also support time transfer over a 30-m free-space link.

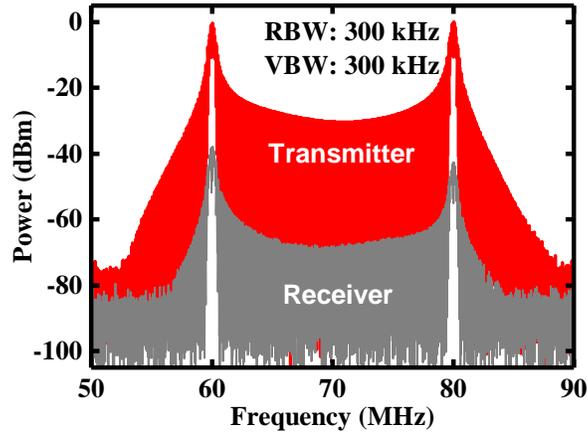

**Fig. 6.** Typical RF spectra of the FlexBOC signals in the transmitter and receiver.

The time delay fluctuations of one-way time interval are depicted in Fig. 7(a) and 7(b), respectively, and in a high agreement with ambient temperature variation during a 15-hour period. Small asymmetry of the both drifts was probably caused by differential time delay of fibers, cables and temperature-sensitive devices. Because of environmental turbulence across the free-space link, the peak-to-peak drifts in one way were ~170 ps. According to two-way transfer methodology, this environmental effect could be cancelled out in the assumption of the reciprocity of atmosphere. The peak-to-peak clock time difference was 95 ps, shown in Fig.

7(c). It is clear that the clock time difference still changed following the temperature variation. This is probably caused by differential time delay. The carrier-to-noise ratios were simultaneously recorded and shown in Fig. 7(d). Their variations also coincided with the temperature change, probably resulted from the temperature-sensitive nature of the high-speed diode lasers or the time interval counters in the receivers. We next analyze the performance metrics in terms of Allan deviations in both time and frequency aspects.

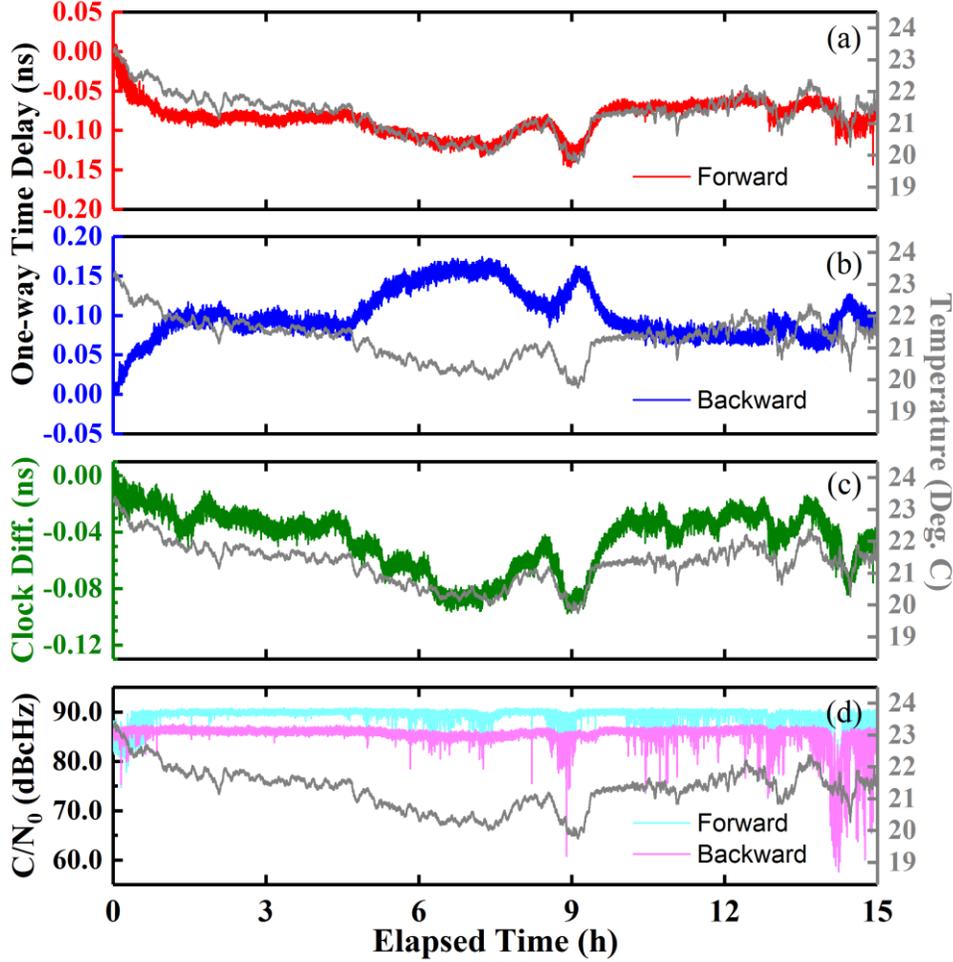

**Fig. 7.** (a) and (b) the time delay fluctuations of one-way time intervals during a 15-h period over a 30 m free-space link in forward and backward directions, respectively. (c) Clock time difference. (d) Carrier-to-noise ratio. Ambient temperature is repeatedly drawn in each subplot for clear view. Clock diff, Clock difference; C/N0, carrier-to-noise ratio.

The fractional frequency instability is given by the modified Allan deviation and exhibited in Fig. 8(a) for the data in Fig.7(c). It reaches $4.0 \times 10^{-12}$ at a gate time of 1 s, which is close to the performance of the frequency reference, and averages down as $\tau^{-1}$ within ~60 s. At 10000 s, the frequency stability approaches to $2.6 \times 10^{-15}$. The time transfer stability is shown in Fig. 8(b) for the same data in Fig. 8(a). The time deviation within a 1-s averaging time is 2.3 ps. It drops to 1.0 ps until ~60 s and then reaches a floor, which approaches to the estimated limit 29. We could almost attribute the time deviation drift beyond an average time of ~60 s to the differential time delay, which could be further reduced by active temperature control of the fibers, cables and transceivers.

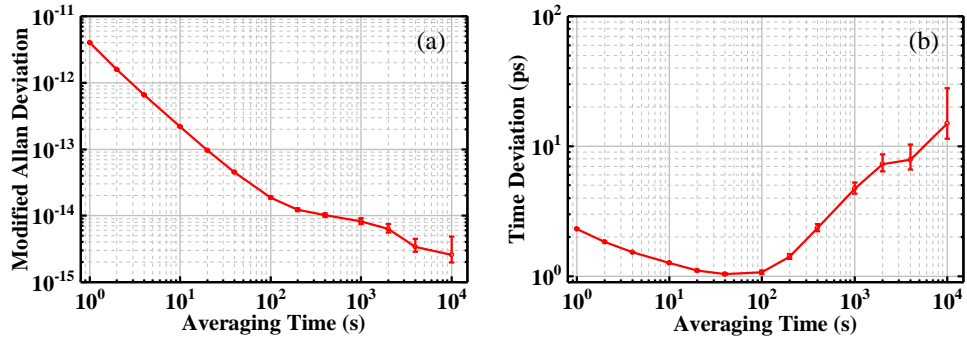

**Fig. 8.** (a) Modified Allan deviation and (b) Time deviation over a 30-m free-space path.

## 5. Discussion

As other two-way time and frequency transfer techniques, our method also takes advantages of the reciprocity of the link. The factors of the reciprocity have been clearly explained in several literatures [24,31], and are valid in our method as well.

In our current system, we utilized a Rubidium frequency standard as the common reference. Therefore, the signal synthesis in the transmitter and the test units in the receiver are limited by the performance of the frequency standard. Right now, the performance of time and frequency transfer is close to the Rubidium frequency reference. According to Reference [30], the time transfer stability at such a high carrier-to-noise ratio, shown in Fig. 5(d), is limited to ~0.8 ps. A better performance would be expected if a more stable frequency standard was employed, such as a Hydrogen maser or photonic microwave divided down from a state-of-the-art cavity-stabilized laser.

Optical terminals and beam steering are also crucial when the traversing range is scaled up to several kilometers, or even ground-to-satellite distance [32]. In our demonstration, two commercialized fiber collimators were utilized, and the insert losses were ~10 dB since the path length of 30 m is far beyond the nominal working distance (~6 m). A pair of telescope with a larger diameter is essential to decrease beam divergence, and thus to enlarge the receiving cross-section for high power-receiving efficiency. In addition, beam misalignment caused by optomechanics creep and beam wander due to environmental variation could lead to a drastic fluctuation of carrier-to-noise ratio that is obvious in the end of the time transfer in Fig. 5(d), or even ruin the link. This could be real-time compensated by steering tip-tilt mirrors.

In our system, the transfer link could be recovered from active jamming due to inner re-acquisition and tracking mechanism of FlexBOC signal. The incoherent detection relaxes the requirements of atmospheric coherence and laser coherence, which matter in coherent detection [24]. Therefore, low-cost, broad-linewidth diode lasers are allowed in this system. Indeed, this results in a degradation of precision compared to coherent detection.

Next, the precision of our system will be enhanced by carrier-phase technique [5,7]. High-speed modulation is needed to fully exploit the advanced technique, and could be realized by frequency upconverter and downconverter. Right now, the optical modules support an S-band modulation, which is capable to a 40~50-folded enhancement referred to the present 70 MHz baseband if perfect frequency synthesis is assumed. For higher precision, external Ku- or Ka-band modulation, i.e., by electro-optical modulation, is intriguingly adopted, and more attention is should be paid to on low-noise frequency converter and O/E converter.

At last, the combination of optical technology and BOC-modulated time and frequency transfer would push forward high-precision clock networks. Currently, the BOC-modulated time and frequency transfer in microwave band has been applied in navigation satellite systems, such as GPS and BeiDou, due to its narrow bandwidth occupation and flexible expansibility [27]. However, multi-path effect of microwave reflection on the multiple surfaces of the devices adjacent to the ground-to-satellite link blurs the receiving signals, and thus remarkably contributes to the uncertainty when a picosecond-precision time transfer is concerned [33]. Because of the excellent directivity, the adoption of laser light as transmission carrier can eliminate the effect. The proposed method might show a potential solution of near-future improvement of navigation satellite systems, and promote ground-to-satellite/inter-satellite optical clock networks.

## 6. Conclusion

We have demonstrated a free-space optical time transfer method using FlexBOC modulation. Over a 30-m free-space path outside the laboratory, our home-built system yields a time deviation of 2.3 ps at a gate time of 1 s, and averages down to 1.0 ps until ~60 s. The fractional frequency instability is $4.0 \times 10^{-12}$ at 1 s, and approaches to $2.6 \times 10^{-15}$ at 10000 s. As discussed above, this combinative approach could be agilely embedded into the current microwave satellite time and frequency transfer system, and avoid the multi-path effect due to microwave reflection. What's more, the performance could be further enhanced by carrier-phase technique. This method should be applicable to near-future ground-to-satellite/inter-satellite clock networks and outdoor timing services.

### Funding

This work was partly supported by the Open Research Fund of the State Key Laboratory of Precision Measurement Technology and Instruments (Grant No. DL18-02).

### Disclosures

The authors declare no conflicts of interest.